\documentclass[twocolumn,prl,english]{revtex4-1}
\usepackage[T1]{fontenc}
\usepackage[latin9]{inputenc}
\usepackage{verbatim}
\usepackage{amsmath}
\usepackage{amssymb}
\usepackage{graphicx}
\usepackage{babel}

\makeatletter

\newcommand{\vecr}{\mathbf{r}}
\newcommand{\vecp}{\mathbf{p}}
\newcommand{\vecn}{\mathbf{n}}
\newcommand{\vecv}{\mathbf{v}}
\newcommand{\vech}{\mathbf{h}}
\newcommand{\vecx}{\mathbf{x}}
\newcommand{\vecnu}{\boldsymbol{\nu}}
\newcommand{\matrixomega}{{{\boldsymbol{\omega}}}}
\newcommand{\matrixu}{{{\mathbf{u}}}}
\newcommand{\matrixsigma}{{{\boldsymbol{\sigma}}}}

\newcommand{\matrixQ}{{{\boldsymbol{Q}}}}
\newcommand{\matrixI}{{{{\mathbf{I}}}}}
\newcommand{\D}{\mathrm{d}}

\makeatother

\begin{document}

\title{Shear-induced first-order transition in polar liquid crystals}

\author{Tomer Markovich$^{1,2}$, Elsen Tjhung$^{1}$ and Michael E. Cates$^{1}$}
\affiliation{
$^{1}$DAMTP, Centre for Mathematical Sciences, University of Cambridge, Wilberforce Road, Cambridge CB3 0WA, United Kingdom \\
$^{2}$Center for Theoretical Biological Physics, Rice University, Houston, TX 77030, USA}
\email{tm36@rice.edu}

%
%
%
%
%

\date{\today}

\begin{abstract}
The hydrodynamic theory of polar liquid crystals is widely used to describe biological active fluids as well as passive molecular materials.  
Depending on the `shear-alignment parameter', 
in passive or weakly active polar fluids under external shear the polar order parameter $\vecp$ is either inclined to the flow at a fixed  (Leslie) angle, or rotates continuously. 
Here we study the role of an additional `shear-elongation parameter' that has been neglected in the recent literature and causes $|\vecp|$ to change under flow.
We show that this effect can give rise to a shear-induced first order phase transition 
from isotropic to polar, and significantly change the rheological properties of both active and passive polar fluids.
\end{abstract}
\maketitle


Liquid crystals typically comprise rod-shaped particles that can align by spontaneously breaking rotational symmetry~\cite{deGennes}. 
Active liquid crystals, in particular, are an important class of non-equilibrium systems where chemical fuel is converted into motion locally~\cite{Marchetti13}.
For example, in actomyosin networks, myosin motors pull the actin filaments  creating  a lengthwise contractile motion.
Self-propulsion in bacterial suspensions has the opposite (extensile) effect.
Crucially, ordering in these biological active fluids can be \emph{polar} rather than nematic: at mesoscopic scales there is not merely an orientational axis but a preferred up/down direction along it.

Polar order in passive, molecular liquid crystals is relatively rare and may originate from electromagnetic interactions~\cite{Ma18,Mertelj17,Biscarini91}.
It has been realized in simulations of interacting dipolar spheres~\cite{Wang02,Patey92} 
and seen under narrow experimental conditions~\cite{Albrecht97}.
Recently~\cite{Mertelj13}, it was found that suspensions of magnetic platelets in nematic solvents can also behave like polar liquid crystals at room temperature. 
To understand their properties, and those of the active systems mentioned above,  a well-founded hydrodynamic theory of polar liquid crystals is needed. 
We argue below that a conventional approach, which effectively assumes fixed magnitude of the polar order, is insufficient. 
We offer an improved description that predicts significantly changed behavior.

Polar liquid crystals contain particles (labelled $i$) whose orientation defines a unit vector ${\vecnu}_i$, pointing from tail to head.
In molecular liquid crystals, polarity is associated with an electric/magnetic dipole moment,
whereas in bacteria and actin filaments, it is viewed as structural.
In (uniaxial) nematic phases, ${\vecnu}_i$ tends to align parallel or anti-parallel along some director field ${\vecn}(\vecr,t)$. 
Here ${\vecn}$ is a ``headless'' unit vector defining the major axis of the mesoscopic tensor $\matrixQ = \langle\vecnu_i\vecnu_i\rangle_{\rm meso}-\matrixI/d = S (\vecn\vecn-\matrixI/d)$. 
(Here $\matrixI$ is the unit tensor and $d$ the dimensionality.) In $d = 3$ the isotropic-nematic transition is first order so that $S$ jumps discontinuously from zero to a finite value~\cite{deGennes}. 
Within the nematic, further changes in $S$ are often neglected, and the dynamics of ${\vecn}$ is well described by the Leslie-Ericksen theory~\cite{deGennes}. 
This contains a shear-alignment parameter, $\xi_0$, which controls whether $\vecn$ reaches steady state at a finite angle to the flow axis (Leslie angle) or rotates indefinitely in time. 

On the other hand, in polar materials, ${\vecnu}_i$ tends to point along the polarization field $\vecp(\vecr,t)$. 
This is a true vector, defined as $\vecp = \langle\vecnu_i\rangle_{\rm meso}$.
In contrast to nematics, the isotropic-polar transition is generically continuous~\cite{deGennes}. 
This allows access to weakly ordered liquid crystals in which the magnitude of $\vecp$ is highly susceptible to perturbations, including flow.
Despite this, in much of the literature~\cite{Marchetti13, Furthauer12, Kruse2004, Kruse05, Giomi10, Tjhung13, Tjhung13b,Cates17,Joanny07,Giomi08,Loisy18}, 
the dynamics of $\vecp(\vecr,t)$ is effectively described by the Leslie-Ericksen theory, 
in which $|\vecp|$ is not changed by flow.

We argue in this Letter that, because $\vecp$ is not a unit vector, 
its dynamics under shear should include an extra term, governed by a {\it shear-elongation parameter,} $\xi_2$. 
Under flow this can increase (or in some cases, decrease~\cite{supp}) the magnitude of $\vecp$:  particles become more aligned with shearing.
This can give rise to interesting unforeseen physics, specifically a shear-induced first order isotropic-polar transition, in both the passive and active cases. 
We will show this analytically for free-anchoring conditions on $\vecp$ at the system boundaries, confirming it numerically under broader conditions.

\begin{figure}
\begin{centering}
\includegraphics[width=1.0\columnwidth]{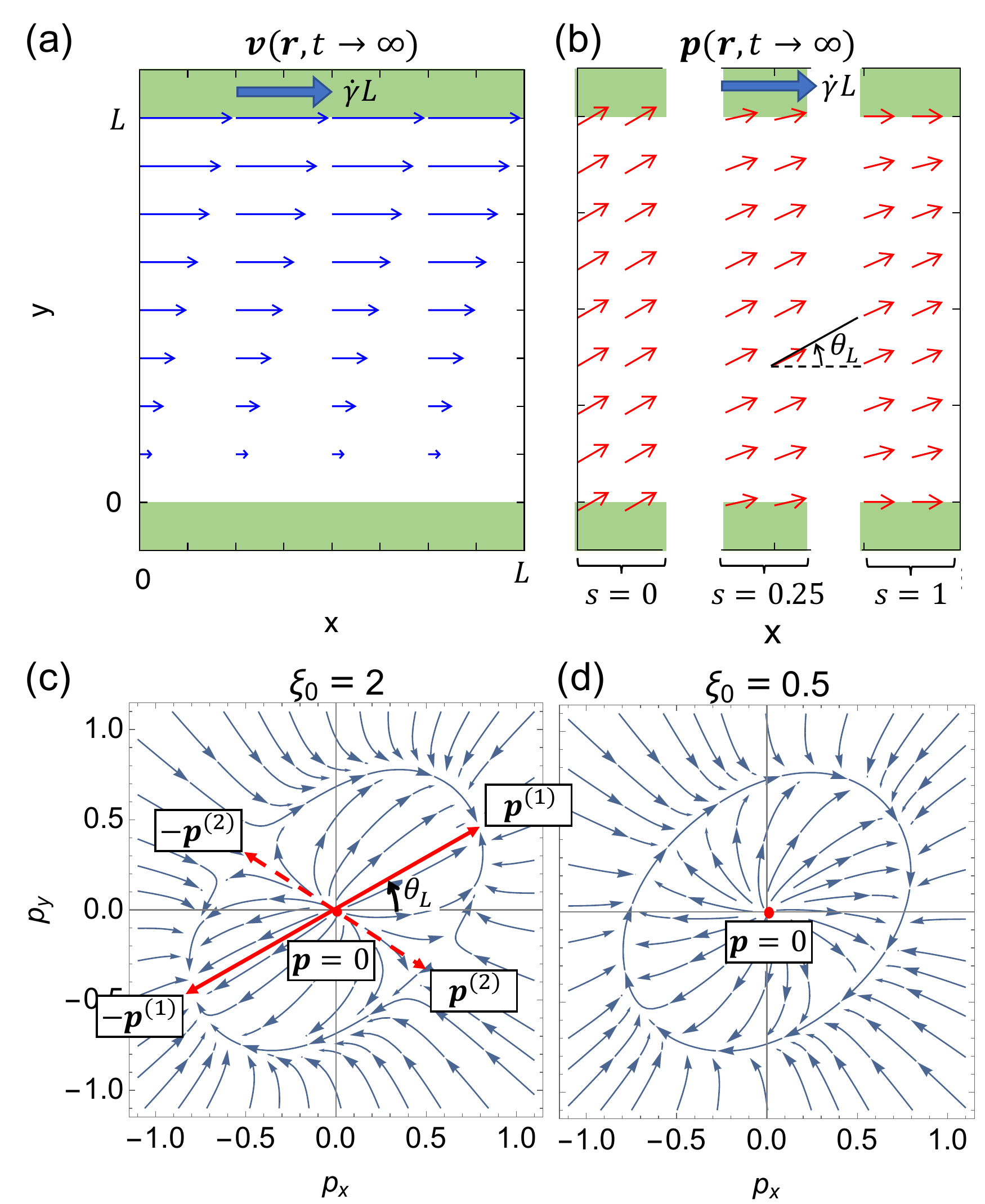}
\par\end{centering}
\caption{
(a) The geometry considered in our model: the upper plate is sheared with fixed velocity, $\dot\gamma L$, to the right while the lower plate is kept stationary.
(b) For $\xi_0>1$, at steady state, $\vecp$ aligns with the shear flow with positive Leslie angle, $\theta_L$. 
This panel shows the steady state $\vecp$ for increasing anchoring strength $s$ from left to right.
(c) Stability flow of $\vecp(t)$ for $\xi_0>1$ (see equations (\ref{eq:px}-\ref{eq:py})). The stable fixed points are $\vecp^{(1)}$ and $-\vecp^{(1)}$.
(d) For $|\xi_0|<1$ we get a limit cycle, which corresponds to the shear-tumbling regime.}
\label{fig:snapshots}
\end{figure}


Our hydrodynamic variables are the order parameter, $\vecp(\vecr,t)$, and the fluid velocity, $\vecv(\vecr,t)$.
As usual we assume $\nabla\cdot\vecv=0$, so that the density, $\rho(\vecr,t)$, is constant.
%
%
The free energy is $F[\vecv,\vecp] = \int \left( \frac{1}{2}\rho v^2 +  f_0 \right) \, \D V$,
whose non-kinetic part $f_0$ can be written in Landau-Ginzburg form:
\begin{equation}
f_0 = \frac{A}{2}|\vecp|^2 + \frac{B}{4}|\vecp|^4 + \frac{C}{6}|\vecp|^6 + \frac{K}{2}(\partial_\alpha p_\beta)^2 \, , \label{eq:F}
\end{equation}
with $B$, $C$, and $K$ positive.
$K$ is the elastic constant associated with splay, bend and twist deformation of $\vecp$, 
in the one-constant approximation~\cite{deGennes}.
The temperature-like parameter, $A$, controls the isotropic-polar transition.
Generically one assumes $B>0$,  so that this transition happens continuously at a critical $A =A_c$, where $A_c= 0$ at mean-field level.
For $B>0$ the $C$-term is redundant in the absence of flow. However, as we shall see later, $B$ can be driven negative by strong shear, so we must retain this term to maintain dynamic stability.

The dynamics of $\vecp(\vecr,t)$ can be written as~\cite{Kruse05,Cates17,Tomer-renormalized-viscosity}:
\begin{equation}
\frac{\partial p_\alpha}{\partial t} + (\vecv\cdot\nabla) p_\alpha = -\omega_{\alpha\beta}p_\beta 
														   + \xi_{\alpha\beta}(\vecp)u_{\beta\gamma}p_\gamma
														   + \frac{h_\alpha}{\Gamma} \, , \label{eq:p}
\end{equation}
where $u_{\alpha\beta}$ and $\omega_{\alpha\beta}$ are the symmetrized and anti-symmetrized velocity gradient tensors.
The left hand side in (\ref{eq:p}) includes advection of $\vecp$ by the fluid velocity, $\vecv$.
The first term on the right hand side is the rotation of $\vecp$ by the fluid vorticity, $\matrixomega$.
The second term, $\xi_{\alpha\beta}$, is a generalized alignment term discussed below.
The third term, involving the molecular field, $\vech = -\delta F/\delta \vecp$, serves to relax $\vecp$ towards its free energy minimum, with 
$\Gamma>0$  the rotational friction constant.
Note that the coefficients of $(\vecv\cdot\nabla)\vecp$ and $-\matrixomega\cdot\vecp$ are both unity as required by Galilean and rotational invariance.

The coefficient of the $\matrixu\cdot\vecp$ term in (\ref{eq:p}) is a second rank tensor that depends on $\vecp$.
Its most general form is:
\begin{eqnarray}
\nonumber \xi_{\alpha\beta}(\vecp) &=& \xi_0 \delta_{\alpha\beta} + \xi_2 p_\alpha p_\beta + \xi_1 \epsilon_{\alpha\beta\gamma}p_\gamma \\
&=& \xi_0 \delta^T_{\alpha\beta} + \left( \xi_0 / |\vecp|^2 +  \xi_2 \right) p_\alpha p_\beta + \xi_1 \epsilon_{\alpha\beta\gamma}p_\gamma\, ,
\label{eq:xi}
\end{eqnarray}
where $\xi_0$, $\xi_1$ and $\xi_2$ are scalars which may depend on $|\vecp|$, 
and $\delta^T_{\alpha\beta} \equiv (\delta_{\alpha\beta}-p_\alpha p_\beta/|\vecp|^2)$ is a transverse projector. 
A special case of (\ref{eq:xi}) was previously derived for a specific microscopic model~\cite{Kung06}.

From now on, we neglect the chiral term $\xi_1$ and assume $\xi_0$ and $\xi_2$ to be constants.
The Leslie-Ericksen shear-alignment parameter, $\xi_0$, determines how
 $\vecp$ aligns ($|\xi_0|>1$) or tumbles ($|\xi_0|<1$) in the shear flow.
Indeed the pure Leslie-Ericksen theory is recovered by setting $\xi_2=-\xi_0/|\vecp|^2$. 
(One also multiplies $\vech$ by $\delta^T_{\alpha\beta}$ in (\ref{eq:p}).)
In this case, the second term of (\ref{eq:xi}) vanishes. However, unlike in nematics where $|\vecn| = 1$, there is no reason for this to be true in polar liquid crystals. 
Relaxing the demand of fixed $|\vecp|$, we release a {\it shear-elongation parameter}, $\xi_2$,  which controls whether shear elongates or compresses $\vecp$.

\begin{figure*}
\begin{centering}
\includegraphics[width=0.9\textwidth]{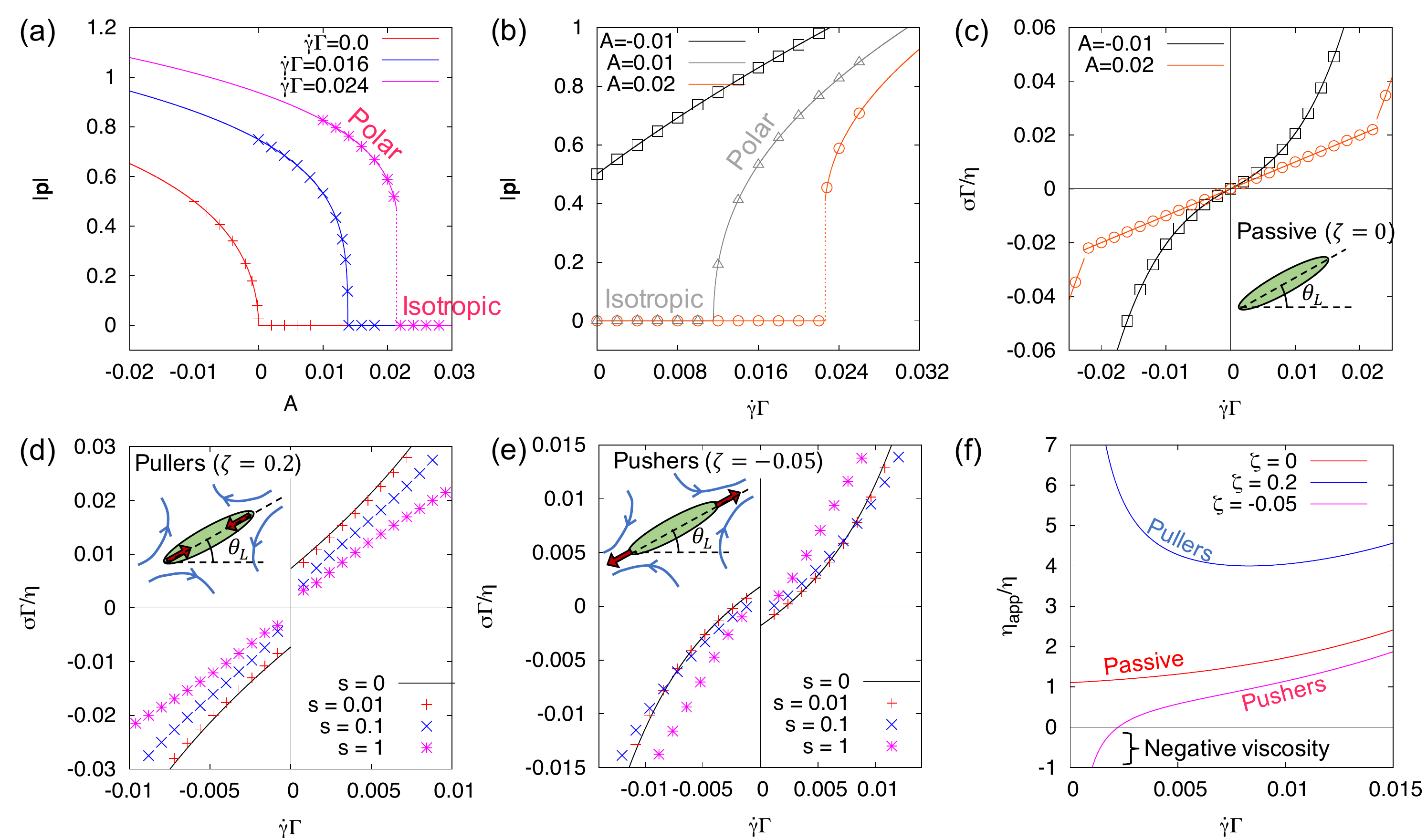}
\par\end{centering}
\caption{
(a) Critical temperature, $A_c$, for polar-to-isotropic transition is shifted upwards with increasing shear rate, $\dot\gamma$. 
Above some threshold, the transition becomes discontinuous.
(b) Generally, particles become more aligned ($|\vecp|$ increases) as we increase the shear rate $\dot\gamma$.
(c) Flow curves for passive polar fluid for different  $A$.
We plot flow curves for contractile (pullers; d) and extensile (pushers; e) cases at various anchoring strength, $s$.
For $s\neq1$ we find a yield stress (positive/negative for pullers/pushers).
(f) Apparent viscosity as a function of dimensionless shear rate, $\dot\gamma\Gamma$, for passive and active polar fluids.
Pushers show a regime of negative apparent viscosity.
Lines are analytic predictions and points are numerical results.
Parameters used: $L=10$, $\rho=2$, $\eta=6.66$, $\dot\gamma=0.0025$, $\Gamma=4$, $A=-0.005$ [in (d,e,f)], 
$B=0.03$, $C=0.04$, $K=0.04$, $\xi_0=2$, and $\xi_2=4$.}
\label{fig:flow-curve}
\end{figure*}

Finally,  $\vecv(\vecr,t)$ obeys the Navier-Stokes equation:
\begin{equation}
\rho\left[ \frac{\partial\vecv}{\partial t} + (\vecv\cdot\nabla)\vecv \right] = -\nabla P + \nabla\cdot\left( \matrixsigma^r + \matrixsigma^d + \matrixsigma^a \right) . \label{eq:v}
\end{equation}
The isotropic pressure $P$ maintains the constraint $\nabla\cdot\vecv=0$.
The dissipative viscous stress is $\matrixsigma^d=2\eta\matrixu$, 
where the shear viscosity, $\eta$, is assumed isotropic.
Like $\vech/\Gamma$ in (\ref{eq:p}) this term dissipates energy as heat, unlike
the reactive stress, $\matrixsigma^r$, whose explicit form is given in~\cite{supp}.
Lastly, $\sigma_{\alpha\beta}^a=\zeta p_\alpha p_\beta$ is a stress arising from activity~\cite{Hatwalne04,self-advection}.
This  is contractile (as for actomyosin) for $\zeta>0$  and extensile (as for bacteria) for $\zeta < 0$; see
Fig.~\ref{fig:flow-curve}(d,e; insets).

The geometry of interest is shown in Fig.~\ref{fig:snapshots}(a,b).
We consider  material confined between parallel plates at $y=0$ and $y=L$
with periodic boundary conditions at $x=0$ and $x=L$.
The upper plate is sheared with  velocity $\vecv(y=L)=\dot\gamma L\hat{\vecx}$ to the right while the lower plate is held stationary: $\vecv(y=0)=0$.
Here $\dot\gamma$ is the average strain rate.
We introduce a variable anchoring strength $s\in[0,1]$ such that
$\partial_y p_x = 0$ and $s p_y = (1-s) \partial_y p_y$ at $y=0$ and $L$.
Strong parallel anchoring at the plates, as often assumed in the literature, is achieved for $s=1$, while $s=0$ indicates free anchoring.
Fig.~\ref{fig:snapshots}(b) shows typical steady state configurations of $\vecp$ for different $s$.

The case of $s=0$ can be solved analytically by
seeking a spatially homogeneous solution for $\vecp$ which satisfy the boundary conditions above.
By symmetry, the fluid velocity, $\vecv$, and pressure, $P$, depend only on the $y$-coordinate.
Therefore $\vecv(y)=\dot\gamma y\hat{\vecx}$ (Couette flow).
Using this we can rewrite (\ref{eq:p}) as:
\begin{eqnarray}
\dot{p}_x &=& \frac{\dot\gamma}{2}p_y(1 + \xi_0 + 2\xi_2 p_x^2) + \frac{h_x}{\Gamma} \label{eq:px} \\
\dot{p}_y &=& \frac{\dot\gamma}{2}p_x(-1 + \xi_0 + 2\xi_2 p_y^2) + \frac{h_y}{\Gamma} \, ,  \label{eq:py}
\end{eqnarray}
while the dynamics in the $z$-direction are irrelevant.

The phase flows of Eqs.~(\ref{eq:px}-\ref{eq:py}) are shown in Fig.~\ref{fig:snapshots}(c,d). For $|\xi_0|>1$,  we find five steady-state solutions:
$\vecp=0$, $\pm\vecp^{(1)}$, and $\pm\vecp^{(2)}$ [Fig.~\ref{fig:snapshots}(c)].
The angle between $\vecp^{(1,2)}$ and the $x$-axis is the usual Leslie angle:
$\theta_L^\pm=\pm\tan^{-1}\sqrt{{(\xi_0-1)}/{(\xi_0+1)}}$.
For $\xi_0>1$, linear stability shows that $\vecp=0$ is unstable, $\pm\vecp^{(2)}$ are saddle points but 
$\pm\vecp^{(1)}$ are stable [Fig.~\ref{fig:snapshots}(c)].
This means $\vecp$ aligns at the positive Leslie angle $\theta_L^+$  [Fig.~\ref{fig:snapshots}(b)].
As $\xi_0\rightarrow1^+$, the Leslie angle approaches zero, and consequently, $\vecp^{(1)}$ and $\vecp^{(2)}$ merge to form a limit cycle.
This is the shear tumbling regime, present for $|\xi_0|<1$, where $\vecp$ rotates clockwise, tracing an ellipse [see Fig.~\ref{fig:snapshots}(d)].
(For nematics, this ellipse becomes the unit circle~\cite{Larson99}.)
Finally for $\xi_0<-1$ (describing, {\it e.g.}, disc-like rather than rod-like particles) $\vecp$ aligns at the negative Leslie angle $\theta_L^-$.

We next study how $|\vecp|$ is affected by shear flow.
In equilibrium ($\dot\gamma=0$ and $\zeta=0$), $|\vecp|$ is given by the minimization of the free energy (\ref{eq:F}).
Under shear, and for $|\xi_0|>1$, 
it turns out that the steady-state $|\vecp|$ is given by the minimization of an \emph{effective} free energy density,
of the same form as the original (\ref{eq:F}),
with renormalized coefficients of $|\vecp|^2$ and $|\vecp|^4$:
\begin{eqnarray}
A &\to& \tilde{A}_{\dot\gamma} = A - \frac{|\dot\gamma|\Gamma}{2}\sqrt{\xi_0^2-1} \label{eq:Atilde} \\
B &\to& \tilde{B}_{\dot\gamma}  = B - \frac{\xi_2}{\xi_0}\frac{|\dot\gamma|\Gamma}{2}\sqrt{\xi_0^2-1} \, . \label{eq:Btilde}
\end{eqnarray}
The coefficient of $|\vecp|^6$ remains unchanged.
As we can see from (\ref{eq:Atilde}), the effect of shear, $\dot\gamma$, is to shift the critical $A_c$ from zero to a higher value.
More interestingly, from (\ref{eq:Btilde}), if $\xi_2/\xi_0>0$, then for large enough $\dot\gamma$, $\tilde{B}_{\dot\gamma}$ becomes negative,
which means the isotropic-to-polar  transition becomes discontinuous.
This is confirmed by numerical simulations of (\ref{eq:p}) and (\ref{eq:v}) [points in Fig.~\ref{fig:flow-curve}(a)].
This contrasts with the isotropic-to-nematic transition under flow which is first-order at rest but disappears via a continuous critical point as $\dot\gamma$ is increased~\cite{Olmsted92}.
In the quasi-Leslie-Ericksen case, $\xi_2=-\xi_0/|\vecp|^2$, $|\vecp|$ minimizes (\ref{eq:F}) at all shear rates because
shear does not stretch or compress $\vecp$. This simplification~\cite{Cates17,Joanny07,Giomi08,Loisy18} therefore misses some important physics of the passive limit.

One can also plot the polarization $|\vecp|$ as a function of shear rate $\dot\gamma$ as shown in Fig.~\ref{fig:flow-curve}(b).
(In molecular liquid crystals, $|\vecp|$ is the net electric/magnetic polarization and can be measured experimentally.)
For $A=0.01$ and $A=0.02$, initially the system is isotropic ($|\vecp|=0$), but as we increase $\dot\gamma$, 
it gains polar order either discontinuously (if $\xi_2/\xi_0>B/A$) or continuously.
For $A=-0.01$, initially the system is polar ($|\vecp|>0$), 
and as we increase $\dot\gamma$, the polar order $|\vecp|$ increases.

Note that in the case of $\xi_2/\xi_0<0$, the sign of $\tilde{B}_{\dot\gamma}$ in (\ref{eq:Btilde}) remains positive and we never get a first-order transition.
Interestingly, in this regime, and in the limit of $|\dot\gamma|\rightarrow\infty$, $|\vecp|$ saturates to a finite value $\sqrt{-\xi_0/\xi_2}$.
Thus, depending on $A$, $|\vecp|$ may decrease with increasing $\dot\gamma$~\cite{supp}.


We can now construct steady-state flow curves for both active and passive cases by the following argument. 
Under homogeneous steady shear,  we apply a shear stress $\sigma$ to the top plate in the $x$-direction [Fig.~\ref{fig:snapshots}(a)].
The rate of work done on the system is then per unit volume
$\dot W_w = \sigma \dot\gamma$.
Similarly, if the liquid crystal is active ($\zeta\neq0$), energy is also continuously injected locally to the system by the active stress $\matrixsigma^a$ at volumetric rate
$\dot W_a= -\sigma_{\alpha\beta}^a u_{\alpha\beta} = -2\zeta p_x p_y u_{xy}$.

By conservation of energy, the total energy injected to the system must match the heat dissipated per unit volume, which is~\cite{Landau86} $\dot Q = \sigma_{\alpha\beta}^d u_{\alpha\beta} + |\vech|^2/\Gamma $.
(Since $\matrixsigma^d=2\eta\matrixu$, $\dot Q$ is positive definite as expected, although
$\dot W_w$ and $\dot W_a$ do not have to be positive.)
Thus, in steady state 
\begin{equation}
\sigma\dot\gamma = \dot W_w =  \dot Q - \dot W_a. \label{eq:Qdot}
\end{equation}
Assuming $\vecp$ to be homogeneous and setting $u_{xy}=u_{yx}=\dot\gamma/2$ in (\ref{eq:Qdot}), 
we obtain the flow curve $\sigma(\dot\gamma)$:
\begin{equation}
\sigma(\dot\gamma) = \left( \eta + \frac{|\vech|^2}{\Gamma\dot\gamma^2} + \zeta\frac{p_x p_y}{\dot\gamma} \right)  \dot\gamma 
            \equiv \eta_\text{app}(\dot\gamma) \dot\gamma,
\label{eq:flow-curve}
\end{equation}
where $p_x=|\vecp|\cos\theta_L$, $p_y=|\vecp|\sin\theta_L$ and $|\vecp|$ is given by 
minimizing the effective free-energy, Eqs.~(\ref{eq:Atilde}-\ref{eq:Btilde}). 
We identify the ratio $\sigma/\dot\gamma$ as the apparent viscosity $\eta_{\text{app}}(\dot\gamma)$.

The result (\ref{eq:flow-curve}) holds for free anchoring ($s=0$). Flow curves for the passive case ($\zeta=0$) are shown in Fig.~\ref{fig:flow-curve}(c) for various $A$.
For $A=-0.01$, the system is already polar at $\dot\gamma=0$, 
and $\eta_\text{app}$ increases monotonically with $\dot\gamma>0$ (shear-thickening).
Note that for this case shear-thinning is also possible~\cite{supp}.
For $A=0.02$, the system is almost isotropic at low strain rate, where $\sigma=\eta\dot{\gamma}$. 
On increasing shear rate, $\sigma(\dot\gamma)$ shows discontinuous shear-thickening
at the shear-induced first order transition, with continuous shear-thickening beyond it. 
Our analytical flow curves fully agree with numerical simulations of (\ref{eq:p}) and (\ref{eq:v}) with free anchoring conditions; see
Fig.~\ref{fig:flow-curve}(c).

We now address the flow curves for active polar fluids ($\zeta\neq0$).
Importantly, the shear-thickening scenarios described above for small $|A|$ remain present in this case, at least for weak activity. 
Deeper in the polar state ($A\ll 0$), local energy injection from the active stress $\matrixsigma^a$ is known to give rise to spontaneously flowing states, 
even in the absence of external shear~\cite{Voituriez05}, when $|\zeta|$ is large enough.
This spontaneous flow transition is mediated by hydrodynamic instabilities in the polarization field $\vecp(\vecr,t)$~\cite{Edwards09,Tjhung13}, 
and (as for nematics~\cite{Cates08,Fielding11}) can give rise to an apparent negative viscosity~\cite{Loisy18,Giomi10}.

Here we consider cases where $|\zeta|$ is below the threshold for spontaneous flow.
In this regime, $\vecp$ remains spatially homogeneous in the bulk fluid (with deviations near the walls if $s\neq 0$). 
At large $\dot\gamma$, activity is negligible and we still find continuous shear-thickening due to shear-elongation. 
In contrast, when $\dot\gamma$ is small, activity is dominant and a yield stress regime is observed (unrelated to spontaneous flow~\cite{Lopez15,Loisy18}).
Strong anchoring ($s \simeq 1$) suppresses this yield-stress regime, but only in small enough systems (see~\cite{supp}). 

Flow curves for pullers ($\zeta>0$) are shown in Fig.~\ref{fig:flow-curve}(d),
where the solid line is (\ref{eq:flow-curve}), and the points are numerics for $s>0$. Due to the opposite effects of activity (thinning) and shear-elongation (thickening), 
the apparent viscosity is non-monotonic, Fig.~\ref{fig:flow-curve}(f).
%
%
Under flow, the system adopts the positive Leslie angle, $\theta_L$ (Fig.~\ref{fig:flow-curve}(d), inset).
Both $p_x$ and $p_y$ are then positive, so that for pullers $\dot{W}_a<0$: activity seemingly removes energy instead of injecting it. 
A detailed description of the flow around each puller would resolve this anomaly: the work done at the walls increases because {\em local} strain rates exceed $\dot\gamma$~\cite{Shelley13}. 
Moreover, allowing for the local energy budget including fuel consumption, the active heat production remains positive~\cite{TOMER}.

Finally, flow curve for pushers ($\zeta<0$) are shown in Fig.~\ref{fig:flow-curve}(e).
Here we observe a negative yield stress and consequently a regime of negative viscosity at small shear rate.
In this regime, $\dot{W}_a>\dot Q>0$ but $\dot{W}_w<0$ from (\ref{eq:Qdot}).
This means the particles are doing macroscopic work on the wall.
Fig.~\ref{fig:flow-curve}(f) shows the apparent viscosity curves $\eta_\text{app}(\dot\gamma)$ corresponding to the flow curves shown in Fig.~\ref{fig:flow-curve}(c,d,e).

Many of the above results for the rheology of active polar liquid crystals echo what was known previously from numerics on active nematics~\cite{Cates08,Fielding11}. 
However it is remarkable that, in the polar case with free-anchoring conditions, the entire scenario is analytically solvable via (\ref{eq:flow-curve}).
More importantly, as stated previously, active polar systems such as bacterial suspensions and actomyosin gels 
can undergo the same shear-induced first order transition as we found for the passive case. 
This could transform their flow behavior with, {\it e.g.}, discontinuous shear-thickening in a system of pushers whose active stress would lead one to expect only a viscosity reduction.

In conclusion, we have presented a hydrodynamic description of polar liquid crystals that generalizes earlier treatments significantly.
In particular, we studied the role of a {\it shear-elongation parameter}, $\xi_2$, 
and showed that this can cause shear-induced first-order transitions from the isotropic to the polar phase in active and passive materials.
In the active case we also found that $\xi_2$ allows non-monotonic apparent viscosities $\eta_\text{app}(\dot\gamma)$, 
and that phenomena such as positive and negative yield stresses, and negative apparent viscosity (resembling those for active nematics) are possible even for weak activity and strong anchoring. 
Finally, we outlined the thermodynamic energy/heat budgets in these different regimes and will return to this in future work~\cite{TOMER}.

\begin{acknowledgments} 
\emph{Acknowledgments:} Work funded in part by the European Research Council under the EU's Horizon 2020 Programme, grant number 760769. 
TM acknowledges funding from the Blavatnik postdoctoral fellowship programme and the National Science Foundation Center for Theoretical Biological
Physics (Grant PHY-1427654). MEC is funded by the Royal Society.
\end{acknowledgments}



\begin{thebibliography}{99}
\bibitem{self-advection}
Additionally, active particles can also swim in the direction of $\vecp$, adding 
a term $\propto (\vecp\cdot\nabla)p_\alpha$ in the left hand side of (\ref{eq:p}). However this does not affect our results which are for homogeneous systems
\end{thebibliography}
\end{document}


\title{Supplementary Material \\ Shear-induced first-order transition in active and passive polar liquid crystals}

\author{Tomer Markovich$^{1,2}$, Elsen Tjhung$^{1}$ and Michael E. Cates$^{1}$}
\affiliation{
	$^{1}$DAMTP, Centre for Mathematical Sciences, University of Cambridge, Wilberforce Road, Cambridge CB3 0WA, United Kingdom \\
	$^{2}$Center for Theoretical Biological Physics, Rice University, Houston, TX 77030, USA}
\email{tm36@rice.edu}

\maketitle

\vspace{-1cm}

\section{Explicit form of the reversible stress}

The reversible stress $\matrixsigma^r$ can be derived by the methods described in~\cite{Cates17}:
\begin{eqnarray}
\sigma^r_{\alpha\beta} &=& (f + \vecp\cdot\vech)\delta_{\alpha\beta} - \frac{\partial f}{\partial(\partial_\beta p_\gamma)}(\partial_\alpha p_\gamma) \nonumber\\
					  &+& \frac{1}{2}(p_\alpha h_\beta - p_\beta h_\alpha) - \frac{1}{2} (p_\alpha\xi_{\beta\gamma}h_\gamma + p_\beta\xi_{\alpha\gamma}h_\gamma),
					  \label{eq:sigmar}
\end{eqnarray}
where $f$ is the free energy density: $F=\int f\,dV$.
Note that the reversible stress is defined together with its corresponding Gibbs-Duhem relation, 
\begin{equation}
\partial_\beta\sigma^r_{\alpha\beta}-p_\beta\partial_\alpha h_\beta = 0.
\end{equation}
A different (equivalent) definition of $\matrixsigma^r$ and a corresponding Gibbs-Duhem relation can be found in~\cite{Joanny07}.

\section{Comparison to other active polar liquid crystal theories}

In much of the active polar liquid crystal literature(\emph{e.g.},~\cite{Loisy18,Giomi08}), the magnitude of $\vecp(\vecr,t)$ is effectively kept constant under shear. 
However, as discussed in the main text, in reality, rodlike particles typically tend to align more in the presence of an external shear.
In other words, $|\vecp|$ should increase with increasing shear rate, $|\dot\gamma|$.
We capture this phenomenology by introducing a shear-elongation parameter, $\xi_2$.
Since $|\vecp|$ can now increase with $|\dot\gamma|$, the rheological properties of active polar liquid crystals can be different as compared to $|\vecp|=$ constant.

\begin{figure}[h]
\begin{centering}
\includegraphics[width=0.6\columnwidth]{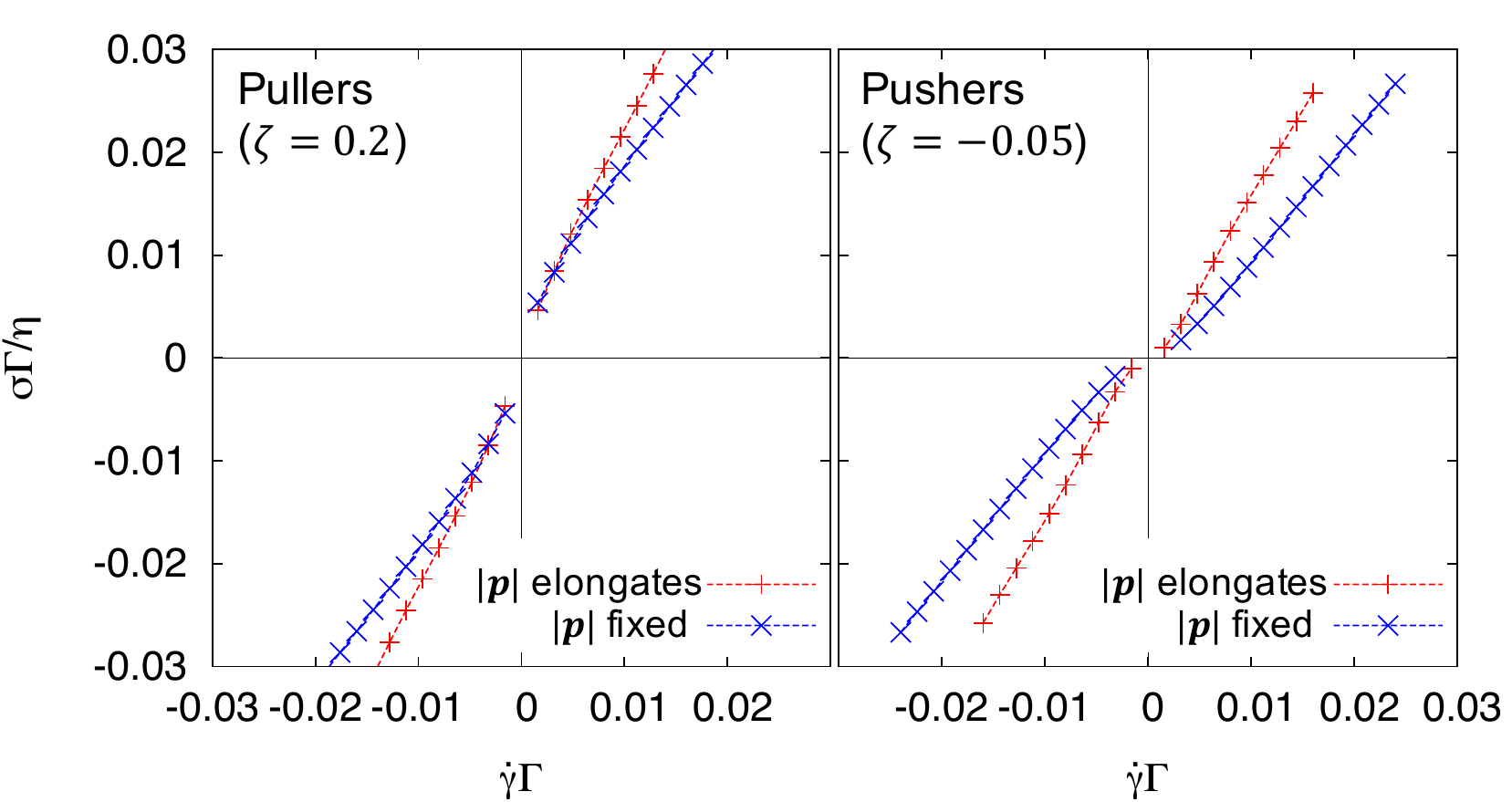}
\par\end{centering}
\caption{
Flow curves for pullers (left) and pushers (right) in the case of strong parallel anchoring at the walls ($s=1$).
Red: Our model with $\xi_0=2$ and $\xi_2=4$, which elongates $|\vecp|$ with increasing $|\dot\gamma|$.
Blue: Leslie-Ericksen limit of our model: $\xi_0=-\xi_2|\vecp|^2=2$, in which case, $|\vecp|=$ constant.}
\label{fig:fixed-p}
\end{figure}

As explained in the main text, the steady-state value  of $|\vecp|$ can be fixed constant (independent of shear rate) by choosing the shear elongation parameter to be:
\begin{equation}
\xi_2 = -\frac{\xi_0}{|\vecp|^2}. \label{eq:constant-p}
\end{equation}
In this case, the constant value of $|\vecp|$ is given by the minimum of the equilibrium free energy [equation (1) of the main text] at steady state.
Alternatively, $|\vecp|$ can also be fixed constant by introducing a Lagrange multiplier to the molecular field, $\vech$, as done in~\cite{Loisy18}:
\begin{eqnarray}
\vech = -\frac{\delta F}{\delta\vecp} + h_\parallel^L \frac{\vecp}{|\vecp|}, \label{eq:constant-p-Lagrange}
\end{eqnarray}
where the value of $h_\parallel^L$ is chosen dynamically (at each point in space and time) so that $|\vecp|=$ constant.
Both (\ref{eq:constant-p}) and (\ref{eq:constant-p-Lagrange}) give the same flow curve. 
We compare the fixed-$|\vecp|$ theory to our theory in Fig.~\ref{fig:fixed-p}.
As we can see from the figure, the shear-elongation parameter can significantly alter the flow curve at large shear rate $|\dot\gamma|$.

\section{System size dependence}

In this section, we investigate the dependence of the flow curve on the system size, $L$, relative to the correlation length, $\ell$.
In the case of weak anchoring conditions, $s\simeq0$, $\vecp$ is spatially homogeneous and aligns with the shear flow at the Leslie angle, $\theta_L^+$.
In this case, the flow curve does not depend on $L$.
However in the case of strong parallel anchoring condition $s\simeq1$, $p_y$ goes to zero at the walls $y=0$ and $y=L$,
but far away from the walls, $\vecp$ again aligns at the Leslie angle.
In between, we expect a spatial variation in $\vecp(y)$ across length scales of order $\ell$.

\begin{figure*}[hb]
	\begin{centering}
		\includegraphics[width=1.0\textwidth]{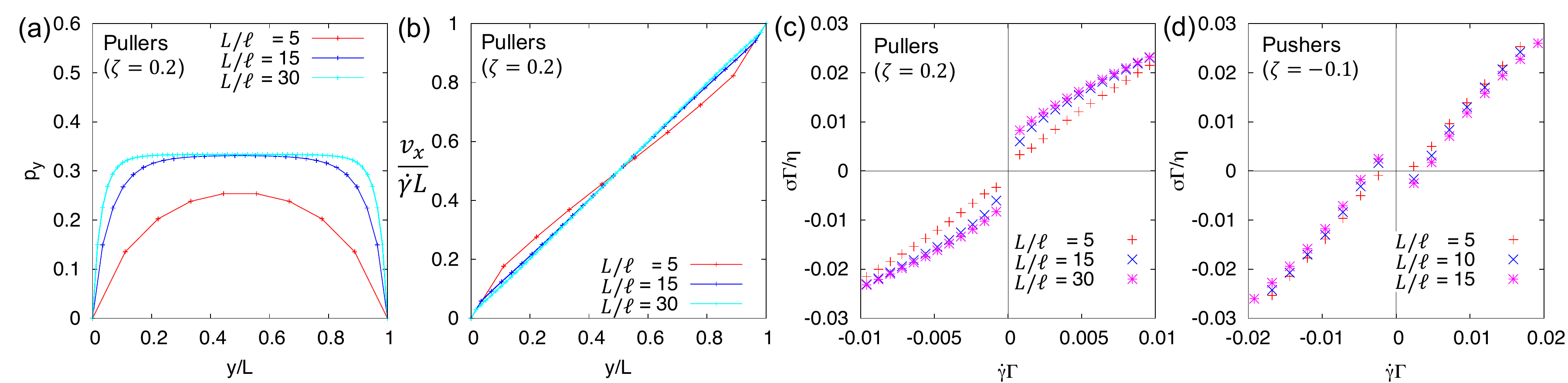}
		\par\end{centering}
	\caption{
		We plot $p_y$ in (a) and the rescaled velocity in (b) as a function of $y/L$ for pullers (plots for pushers are similar). 
		The strong anchoring ($s=1$) imposes $p_y=0$ at the walls as can be seen in (a). 
		One can see in (b) that the fluid velocity is well approximated by the Couette flow $v_x=\dot\gamma y$.
		The flow curves for pullers ($\zeta>0$) and pushers ($\zeta<0$) for different values of $L/\ell$ are presented in (c) and (d), respectively.
		We find that for pullers/pushers a finite positive/negative yield stress emerges as $L/\ell$ increase.
		Parameters used: $s=1$, $\rho=2$, $\eta=6.66$, $\dot\gamma=0.0025$ in (a,b), $\Gamma=4$, $A=-0.005$, $B=0.03$, $C=0.04$,  $\xi_0=2$ and $\xi_2=-4$, $K=0.04$ giving $\ell=1.94$.
	}
	\label{fig:L}
\end{figure*}

From the free energy [equation (1) in the main text], the intrinsic correlation length of the system is given by 
$\ell = \sqrt{-K/(A-C|\vecp_0|^2)}$,
where $|\vecp_0|$ minimizes the free energy (1).
In general, $\ell$ should also depend on the shear rate, $\dot\gamma$, however in this discussion, we take the system to be far from criticality and replace $\ell$ by its value at $\dot\gamma=0$. 
For $L\gg\ell$, we observe a bulk region where $\vecp(y)$ aligns with the Leslie angle [see blue and turquoise lines in Fig.~\ref{fig:L}(a)].
However, when $L$ is comparable to $\ell$, we do not observe this bulk region in $\vecp(y)$ [see red line Fig.~\ref{fig:L}(a)].
From Fig.~\ref{fig:L}(c,d), we see that this bulk region is associated with the appearance of a yield stress in the flow curve. (The yield stress is positive/negative in the case of pullers/pushers.) 
These happen in the weak activity regime, below the threshold for the spontaneous flow transition.
Finally, since we are in the weak activity regime the fluid velocity can be well approximated by the Couette flow $v_x=\dot\gamma y$ [Fig.~\ref{fig:L}(b)].

\section{The case of $\xi_2/\xi_0 < 0$}

\begin{figure*}[h!]
\begin{centering}
\includegraphics[width=0.9\textwidth]{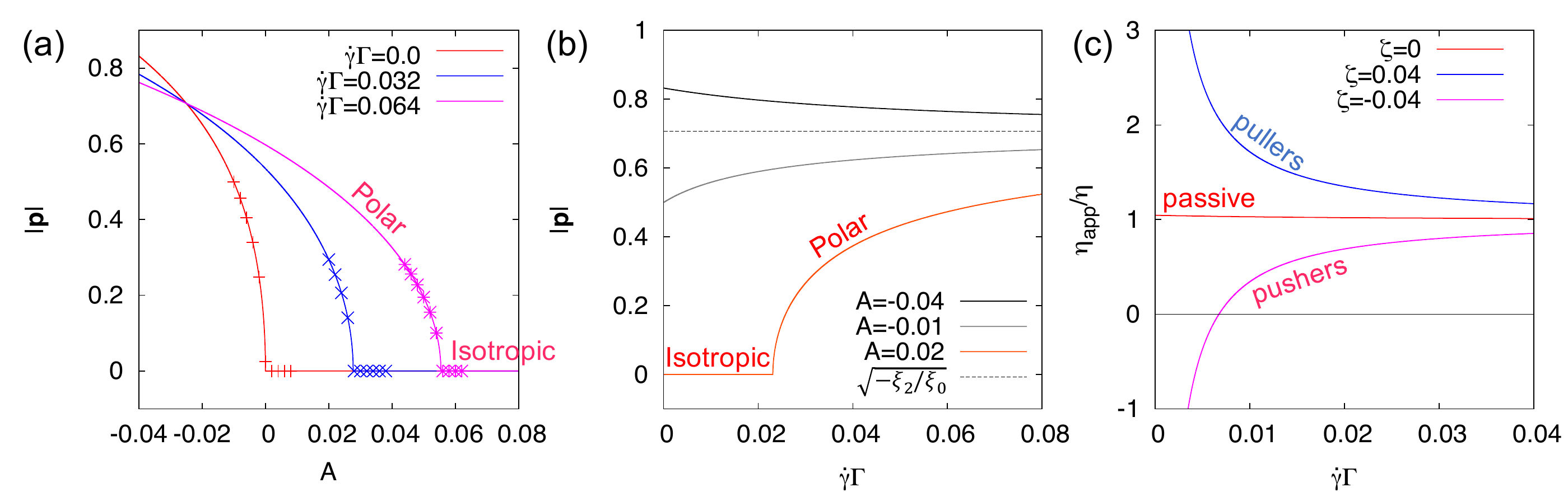}
\par\end{centering}
\caption{
(a) Plot of $|\vecp|$ as a function of rescaled temperature, $A$, for $\xi_2/\xi_0<0$. The transition remains continuous, shear only shifts the critical point, $A_c$, upwards.
(b) Plot of $|\vecp|$ as a function of dimensionless shear rate, $\dot\gamma\Gamma$. $|\vecp|$ approaches $\sqrt{-\xi_2/\xi_0}$ in the limit of $\dot\gamma\rightarrow\infty$.
(c) Apparent viscosity as a function of dimensionless shear rate, $\dot\gamma\Gamma$, for passive and active polar fluid in the case of  $\xi_2/\xi_0<0$
(Lines are theoretical results and points are numerical results).
Parameters used: $L=10$, $s=0$, $\rho=2$, $\eta=6.66$, $\dot\gamma=0.0025$, $\Gamma=4$, $A=-0.03$ (in (c)), $B=0.03$, $C=0.04$, $K=0.04$, $\xi_0=2$ and $\xi_2=-4$.}
\label{fig:negative-xi2}
\end{figure*}

As already mentioned in the main text, in the case of $\xi_2/\xi_0 < 0$, 
$\tilde{B}_{\dot\gamma}$ in the effective free energy remains positive so that the isotropic-to-polar transition remains second-order for all values of $\dot\gamma$
[see Fig.~\ref{fig:negative-xi2}(a)].
Interestingly, depending on the value of $A$, $\vecp$ can either increase or decrease with increasing shear rate $\dot\gamma$ [see Fig.~\ref{fig:negative-xi2}(b)].
The corresponding apparent viscosities are shown in Fig.~\ref{fig:negative-xi2}(c) for both passive and active cases.

